\documentclass[%
 aps,prmaterials,
 amsmath,amssymb,
 reprint,%
]{revtex4-2}

\usepackage{graphicx}
\usepackage{dcolumn}
\usepackage{bm}

\usepackage[utf8]{inputenc}
\usepackage[T1]{fontenc}
\usepackage{mathptmx}
\usepackage{etoolbox}
\usepackage{multirow}

\makeatletter
\def\@email#1#2{%
 \endgroup
 \patchcmd{\titleblock@produce}
  {\frontmatter@RRAPformat}
  {\frontmatter@RRAPformat{\produce@RRAP{*#1\href{mailto:#2}{#2}}}\frontmatter@RRAPformat}
  {}{}
}%
\makeatother
\begin{document}


\title[The structural stability of tungsten nanoparticles]{The structural stability of tungsten nanoparticles}
\author{L. Pizzagalli}
\author{S. Brochard}%
\author{J. Godet}%
\author{J. Durinck}%
 \email{Laurent.Pizzagalli@univ-poitiers.fr}
\affiliation{ 
Institut P' , CNRS UPR 3346, Université de Poitiers, SP2MI, Boulevard Marie et Pierre Curie, TSA 41123, 86073 Poitiers Cedex 9, France
}%

\date{\today}

\begin{abstract}
Motivated by contradicting reports in the literature, we have investigated the structural stability of tungsten nanoparticles using density functional theory calculations. The comparison of BCC, FCC, A15, disordered, and icosahedral configurations unequivocally shows that BCC is the energetically most stable structure when the number of atoms is greater than 40. A disordered structure is more stable for smaller sizes. This result conflicts with an earlier theoretical study on transition metal nanoparticles, based on a semi-empirical modeling of nanoparticles energetics [D. Tom{\'a}nek et al., Phys. Rev. B \textbf{28}, 665 (1983)]. Examining this latter work in the light of our results suggests that an erroneous description of clusters geometry is the source of the discrepancy. Finally, we improve the accuracy of the semi-empirical model proposed in this work, which will be useful to calculate nanoparticle energies for larger sizes.   
\end{abstract}

\maketitle

\section{\label{sec:intro}Introduction}

Due to its high melting and evaporation temperatures and excellent mechanical properties, tungsten is the selected wall material to be used in fusion reactor divertors like ITER. Despite a high thermal stability and a good resistance to sputtering, the interaction with the plasma during operation leads to the formation of dust composed of tungsten nanoparticles of various sizes~\cite{Rub18FED}. This aspect motivated extensive investigations on the properties of W nanoparticles, and in particular their environmental and biological impacts in a fusion-like environment~\cite{Acs20INTECH}. Tungsten-based nanomaterials are also increasingly used in biomedicine applications~\cite{Gu22ADM}. 

The most stable phase of bulk tungsten is named $\alpha$-W and has a body-centered structure (BCC). Two metastable allotropes are known, which can form in specific conditions. The first one is named $\beta$-W and has a A15 cubic structure. $\beta$-W thin films have been extensively studied owing to the report of a giant spin Hall effect~\cite{Pai12APL}. The second one is $\gamma$-W, with a face-centered cubic (FCC) structure~\cite{Cho04APL}. Naturally, one would assume that tungsten nanoparticles are made of the most stable bulk phase, i.e. $\alpha$-W. However, in an early theoretical paper, Tom{\'a}nek and co-workers predicted that a FCC structure would be favored in small nanoparticles for several transition metals, including tungsten~\cite{Tom83PRB}. Their conclusions were based on the argument that surface energies associated with the FCC structure are typically lower than the ones for the BCC structure, and that this effect will dominate at small sizes when the surface-to-bulk ratio increases. In the specific case of tungsten, they calculated a nanoparticle size threshold of 5-6~nm, below which FCC should be the lowest energy structure. Later X-ray diffraction experiments on nanometer sized W clusters confirmed the prediction, and determined the BCC-FCC structural transition at 7~nm~\cite{Oh99JCP}. 
Similar conclusions in support of this argument were reached for Mo and Cr clusters~\cite{Huh00PRB,Vys05APL}. But other experiments led to different observations. For instance Iwama and Hayakawa found that 3--20~nm Mo and W nanoparticles show crystalline structures being either BCC or A15 for Mo, and A15 for W, but not FCC~\cite{Iwa85SS}. More recently, Sch{\"o}ttle and co-workers reported transmission electron microscopy observations of 1-2~nm W nanoparticles with a BCC structure~\cite{Sch14CCOM}.  

On the numerical side, classical molecular dynamics simulations were carried out for investigating this possible BCC-FCC transition as a function of the size and shape in W nanoparticles. Hence Chen and co-workers reported that the BCC structure is significantly more stable than the FCC, except for a singular high energy shape~\cite{Che17JNR}. This work confirms an earlier study by Marville and Andreoni, who found that tungsten  nanoparticles were more stable in a BCC structure than in a FCC or icosahedral arrangement~\cite{Mar87JPCH}. These simulations then support the recent microscopy measurements against the early predictions and experiments. However one has to be cautious about definite conclusions, since interatomic potentials are not always accurate in the description of under-coordinated atoms at the surfaces and edges of nanoparticles. Another aspect is the potential stabilization of the A15 phase in the W nanoparticles, which was overlooked in published numerical works. This motivates us to perform a thorough investigation of the structural stability of small tungsten nanoparticles using first-principles calculations. In particular we consider numerous systems with BCC, FCC, A15, disordered, and icosahedral structures.   

\section{Methods}

\subsection{\label{sec:calc}Electronic structure calculations}

\renewcommand{\arraystretch}{1.3}
\begin{table}
\caption{\label{tab:1}Computed data for tungsten in BCC, FCC, and A15 structures, compared to experimental results when available, and DFT computed data from the literature. The lattice parameter $a_0$ is expressed in angstrom, the bulk modulus $B$ in GPa, and the surface energies $\gamma$ in J~m$^{-2}$. The bulk energy per atom $\varepsilon_0$ are given relatively to the BCC phase (i.e. $\varepsilon_0(\mathrm{BCC})=0$), and are expressed in eV~at$^{-1}$. Note that the energies for \{110\} and \{111\} surfaces of FCC W (marked with a *) correspond to the unrelaxed slab configuration (see text for details).}
\begin{ruledtabular}
\begin{tabular}{llccc}
& & & \multicolumn{2}{c}{DFT} \\\cline{4-5}
& & Exp. & This work & Others \\ \hline
\parbox[t]{2mm}{\multirow{5}{*}{\rotatebox[origin=c]{90}{BCC}}} 
& $a_0$ & 3.165\footnote{Reference~\onlinecite{Las99SPR}} & 3.1753 & 3.1741\footnote{Reference~\onlinecite{Bec06PRL}}, 3.172\footnote{Reference~\onlinecite{Hei10JAP}}\\ 
& $B$ & 310\footnote{Reference~\onlinecite{Low67JAP}} & 311 & 309\footnote{Reference~\onlinecite{Jia20IJQC}}\\ 
& $\gamma_{100}$ & & 4.071 & 3.954\footnote{Reference~\onlinecite{Tra16SD}} \\ 
& $\gamma_{110}$ & 3.265\footnote{Reference~\onlinecite{Tys77SS}}, 3.675\footnote{Reference~\onlinecite{Boe89NOR}} & 3.302 & 3.230\footnotemark[6] \\ 
& $\gamma_{111}$ & & 3.569 & 3.466\footnotemark[6] \\ \hline
\parbox[t]{2mm}{\multirow{6}{*}{\rotatebox[origin=c]{90}{FCC}}} 
& $a_0$ & 4.13\footnotemark[1] &  4.0256 & 4.025\footnotemark[3], 3.960\footnote{Reference~\onlinecite{Ein97PRL}}\\ 
& $\varepsilon_0$ & & 0.494 & 0.49\footnotemark[3], 0.50\footnotemark[9]\\
& $B$ & & 282 & 286\footnote{Reference~\onlinecite{Jon15SD}}  \\ 
& $\gamma_{100}$ & & 3.284 \\ 
& $\gamma_{110}$ & & 3.736$^*$ \\ 
& $\gamma_{111}$ & & 2.637$^*$ \\ \hline
\parbox[t]{2mm}{\multirow{3}{*}{\rotatebox[origin=c]{90}{A15}}}
& $a_0$ & 5.05\footnotemark[1] & 5.0665 & 5.059\footnotemark[3] \\ 
& $\varepsilon_0$ & 0.082\footnote{Reference~\onlinecite{Bar17JCP}} & 0.093 & 0.09\footnotemark[3] \\
& $B$ & & 298 & 298\footnotemark[6]\\
\end{tabular}
\end{ruledtabular}
\end{table}

We perform first principles calculations in the framework of density functional theory (DFT) for determining the relaxed structure and the associated energy of various nanoparticle models. The code \textit{pw.x} in the Quantum Espresso software~\cite{Gia17JPCM} is used for this purpose. A well converged electronic structure is achieved by using a plane wave energy cutoff of 40~Ry and a charge density cutoff of 320~Ry. Exchange-correlation contributions are described with the Perdew-Burke-Ernzerhof functional~\cite{Per96PRL}. We use the Projector Augmented-Wave method~\cite{Blo94PRB} for ion-electron interactions, with the valence electron configuration 5s$^2$5p$^6$5d$^4$6s$^2$. Finally, the cold smearing method~\cite{Mar99PRL} of Marzari and co-workers is applied to improve the convergence of the electronic structure. 

We first compute the lattice parameter and bulk modulus for the three tungsten allotropes, and their energie differences, using supercells and a very dense grid of k-points for Brillouin zone sampling. Our results in Tab.~\ref{tab:1} are in excellent agreement with other recent calculations and available experiments. The table also includes the energies of surfaces with low Miller indexes for BCC and FCC W. Those are calculated using a slab configuration, a large number of layers, and a k-point grid of $20\times20\times1$, in order to obtain converged results. The relaxation is achieved when all components of all ionic forces are lower than $2.6\times10^{-4}$~eV\ \AA$^{-1}$. Calculated surface energies for BCC W are in good agreement with the literature. We also consider the canonical surfaces of FCC W, for which no information seems to be available in the literature. In the case of the $(110)$ and $(111)$ surfaces, we observe distortions occurring in the center of the slab during force relaxation. We have not pursued the analysis of this issue any further, since it is not the focus of the study. For these two surfaces, we report in Tab.~\ref{tab:1} the energies of the unrelaxed initial configuration. 

For the relaxation of nanoparticles, we employ supercells large enough to allow for at least 10~\AA\ between periodic replicas in all dimensions. The k-point sampling is made at the $\Gamma$-point as is usual for 0D systems. Finally, forces are relaxed until all components for all ions are lower than $2.6\times10^{-4}$~eV\ \AA$^{-1}$.

\subsection{\label{sec:models}Nanoparticles selection}

The shape of 1-2~nm nanoparticles, i.e. including about thirty to a few hundred atoms, is often poorly documented, and tungsten is no exception. Therefore we follow a well proven, standard methodology~\cite{Amo21CRP}, in order to generate low energy initial configurations. For BCC and FCC for which the energies of surfaces are known, we use a Wulff construction as a first option. Alternatively, we also carve nanoparticles out of bulk, with spherical or smoothed cubic shapes. A coordination analysis is next performed to select configurations with as few low coordination atoms as possible, since the latter are associated with a high energy in metals. These two techniques are both used to obtain initial configurations of BCC and FCC nanoparticles. For A15 for which surfaces are not known, only spherical nanoparticles are generated. 

The carving strategy is more difficult to implement in the case of a disordered structure, since bulk tungsten does not exist in an amorphous phase to our knowledge. Recently, Jana and Caro performed an extensive search for the most stable structures of iron nanoparticles including less than 200 atoms~\cite{Jan23PRB}. Except for a few crystalline configurations at magic numbers, they found that most of these structures are amorphous. We select several of these configurations from their freely available structures database, which can be used as initial configurations after a small rescaling. These systems are labeled D (for disordered) in the remainder of the paper.   


Finally, we also consider icosahedral nanoparticles, identified with label 'I'. The two possible candidates in the investigated size range contain 55 and 147 atoms~\cite{Bes95JMST}. Overall we select 27 different nanoparticles, with atoms numbers ranging from 55 to 169, to be used as starting configurations in first-principles calculations. The set includes 8 BCC, 7 FCC, 5 A15, 5 disordered (D), and 2 icosahedral (I) nanoparticles.

\section{\label{sec:energy}Stability}

\begin{figure}[th!]
\begin{center}
\includegraphics[width=0.95\linewidth]{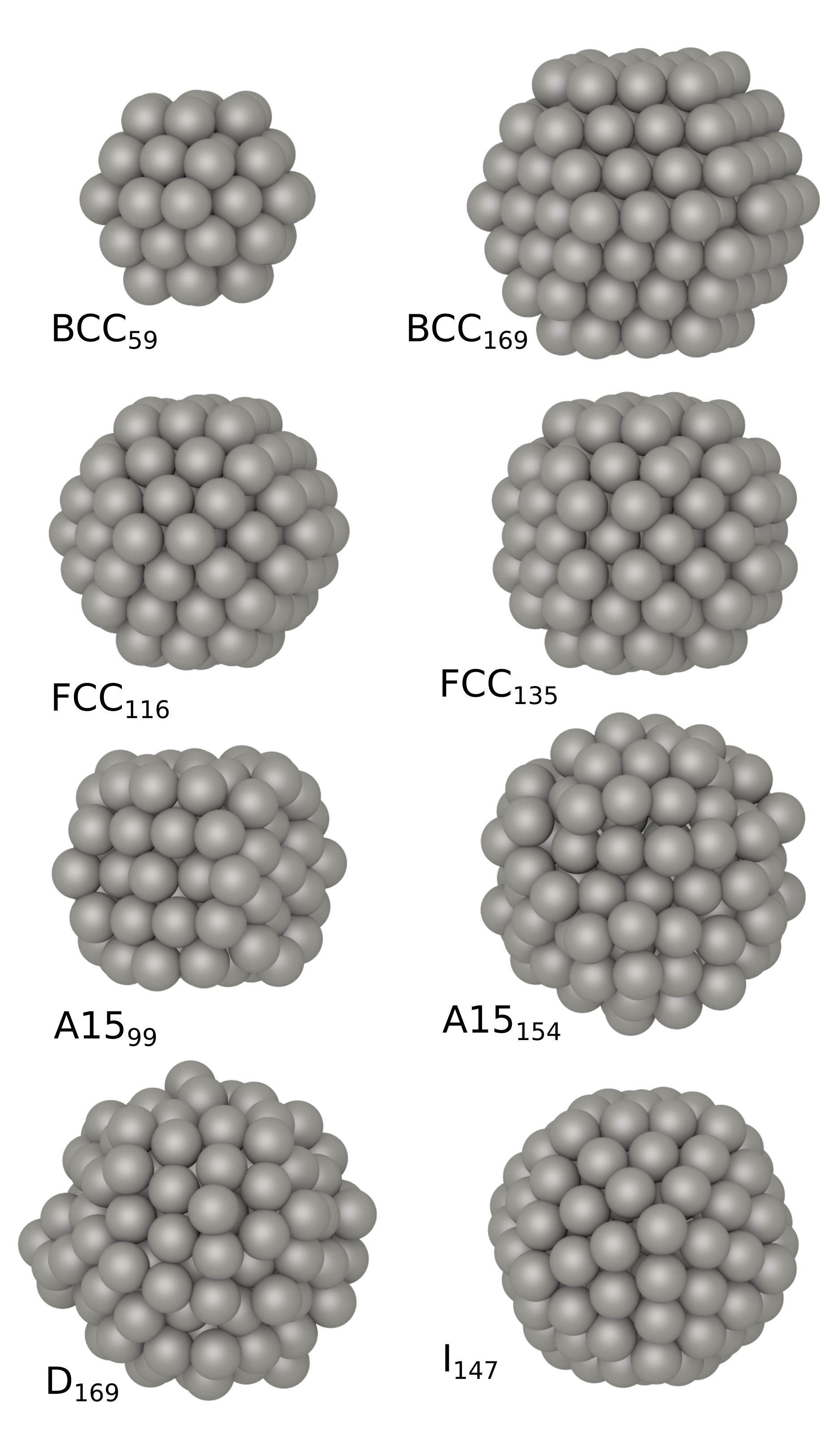}
\end{center}
\caption{\label{fig:1} Examples of DFT relaxed tungsten nanoparticles, with various sizes (given as a number of atom in the label) and different atomic structures (BCC, FCC, A15, D for disordered, I for icosahedral).}
\end{figure}

\begin{figure}[th!]
\begin{center}
\includegraphics[width=0.95\linewidth]{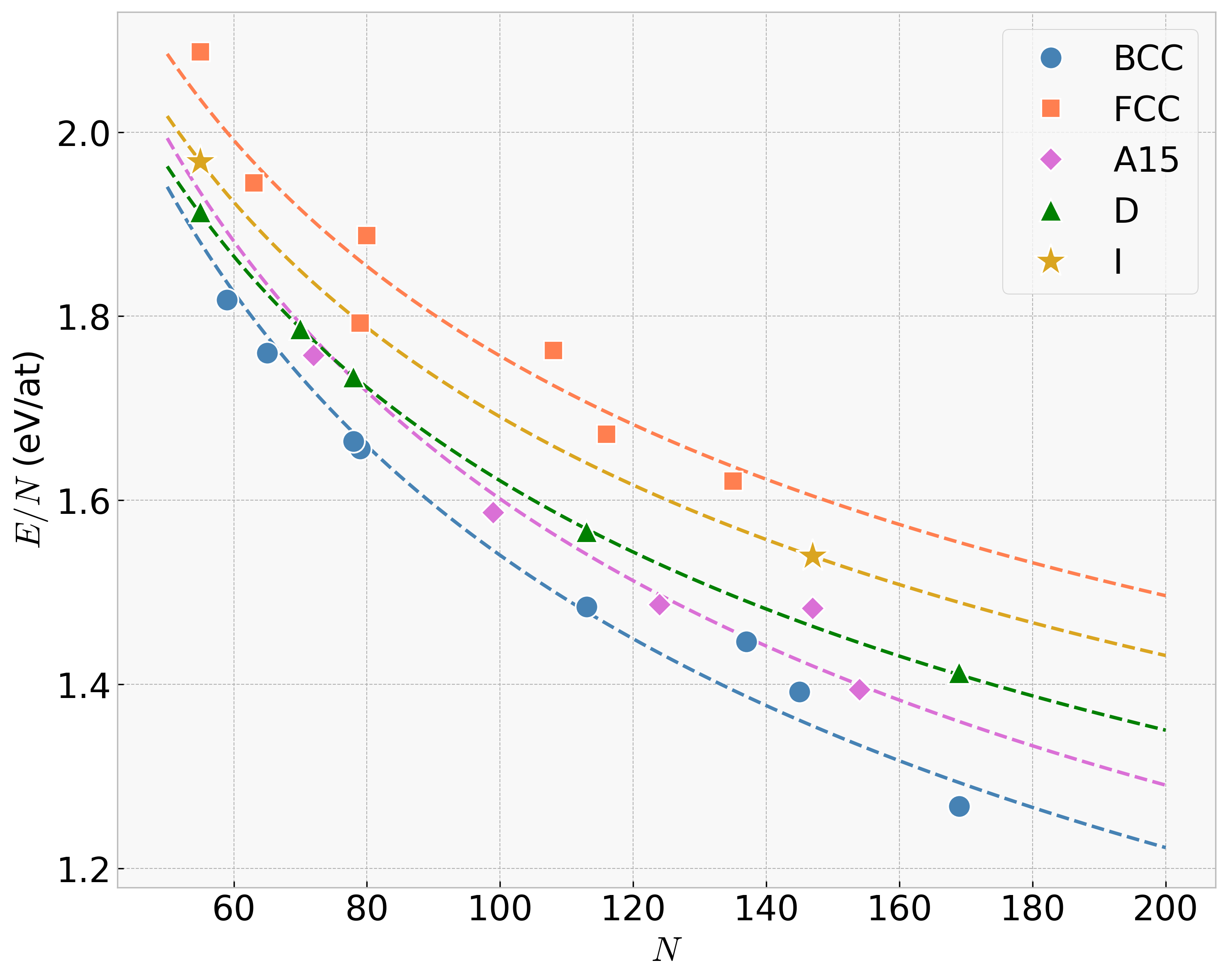}
\end{center}
\caption{\label{fig:2} Nanoparticle energy per atom (eV) as a function of the number of atoms $N$, the reference being the BCC bulk energy, after DFT relaxation (BCC: blue circles; FCC, orange squares; A15: magenta diamonds; D: green triangles; I: golden stars). Dashed lines are obtained by fitting all data points for a given set with Eq.~\ref{eq:3} (fit parameters reported in Tab.~\ref{tab:2}).}
\end{figure}


Selected examples of DFT relaxed nanoparticles are represented in Fig.~\ref{fig:1}. In most cases we observe limited structural changes compared to the initial configurations. In particular only a slight surface relaxation is observed for BCC and I nanoparticles, with an energy gain of 0.15~eV/at in average. For A15 and FCC systems, the surface relaxation is greater, with an average relaxation energy of 0.26~eV/at and 0.32~eV/at, respectively. The larger energy gain is obtained for D nanoparticles, for which significant surface atom displacements are identified. In average, the relaxation of D nanoparticles yields an energy gain of 0.83~eV/at. 

The energy $E$ (with respect to the perfect BCC bulk) of a relaxed nanoparticle made of $N$ atoms can be written

\begin{equation}
E = N\varepsilon_0 + E_s(N), \label{eq:1}
\end{equation}

with $\varepsilon_0$ the bulk energy per atom relatively to the ground state BCC structure ($\varepsilon_0 (\mathrm{BCC})=0$). $E_s$ is an excess energy, akin to a surface energy, although it can in principle includes further contributions associated with Laplace pressure, surface stress relaxation, or quantum confinement effects. Assuming that only surface atoms contributes to $E_s$, and that the number of surface atoms is roughly $N^{2/3}$, Eq.~\ref{eq:1} becomes 
 
\begin{equation}
E = N\varepsilon_0 + \beta N^{2/3}, \label{eq:2}
\end{equation}

with $\beta$ the average energy contribution to $E_s$ from surface atoms. The energy per atom of a nanoparticle  can then be written 

\begin{equation}
E/N = \varepsilon_0 + \beta N^{-1/3}. \label{eq:3}
\end{equation}

%
%

Figure~\ref{fig:2} shows the nanoparticle energies calculated by DFT for all configurations. As predicted by Eq.~\ref{eq:3}, $E/N$ values increase for $N\rightarrow0$ due to the growing surface contribution, and converge to $\varepsilon_0$ for $N\rightarrow\infty$. The most striking result is that for a given $N$, BCC nanoparticles appear to be always more stable than the others. Fitting the Eq.~\ref{eq:3} separately for each kind of structure better highlights this finding (dashed lines in Fig.~\ref{fig:2}). We observe overall $E(\mathrm{BCC})<E(\mathrm{A15})<E(\mathrm{D})<E(\mathrm{I})<E(\mathrm{FCC})$ in the investigated $N$ range. Therefore our results disagree with predictions made in Ref.~\onlinecite{Tom83PRB}. In fact our calculations clearly show that FCC nanoparticles are not energetically more stable than BCC nanoparticles. They also reveal that BCC is the lowest energy structure and that A15, disordered and icosahedral nanoparticles are more stable than FCC nanoparticles. 

\renewcommand{\arraystretch}{1.3}
\begin{table}[th!]
\caption{\label{tab:2}
$\beta$ parameters obtained by fitting Eq.~\ref{eq:3} with DFT calculations (shown in Fig.~\ref{fig:2}). For BCC, FCC, and A15 nanoparticles, $\varepsilon_0$ values calculated for bulk systems are used in the fit. For disordered (D) and icosahedral (I) nanoparticles, $\varepsilon_0$ is computed together with $\beta$ by fit. The table also includes the corresponding surface energies calculated using the spherical model described in the appendix, and the number of atoms $N_c$ below which each phase becomes more stable than BCC.}
\begin{ruledtabular}
\begin{tabular}{lcccc}
& $\beta$ (eV) & $\varepsilon_0$ (eV) & $\gamma$ (J~m$^{-2}$) & $N_c$ \\ \hline
BCC & 7.149 &  0.000 &  3.729 &  \\
FCC & 5.861 &   0.494 &  3.019 & 17 \\ 
A15 & 7.000 &   0.093 &  3.614 &  4 \\
D &  6.099 &  0.307 &  3.142 &  40 \\
I &  5.836 &  0.433 &  3.006 &  27  \\
\end{tabular}
\end{ruledtabular}
\end{table}

The $\beta$ values determined in the fitting process are reported in Tab.~\ref{tab:2}. We adjust $\beta$ while using $\varepsilon_0$ from bulk calculations (Tab.~\ref{tab:1}), except for disordered and icosahedral nanoparticles for which $\varepsilon_0$ is also adjusted. Note that only minor changes are observed when both $\beta$ and $\varepsilon_0$ are adjusted on DFT data for BCC, FCC, and A15 nanoparticles. 

$\beta$ in Eq.~\ref{eq:3} corresponds to an energy per surface atom. Conversion into the usual surface energies $\gamma$, i.e. energies per surface area, is straightforward if one assumes that nanoparticles are spherical (see the Appendix). Computed values are included in Tab.~\ref{tab:2}. We find that the lowest $\gamma$ values are obtained for I and FCC, followed by D, A15, and finally BCC. This is in agreement with arguments based on a lower surface to volume ratio for icosahedral and FCC~\cite{Tom83PRB,Bes95JMST,Huh00PRB}. We also observe that $\gamma$ values for BCC and FCC are in the range of surface energies calculated for well defined orientations as reported in Tab.~\ref{tab:1}. However, it is difficult to push further the comparison. Swaminarayan and co-workers proposed that the surface energy of spherical nanoparticles made of FCC metals is close to $\gamma_{110}$~\cite{Swa94SS}. This is clearly not valid in the present work. Our value of 3.019~J~m$^{-2}$ is also larger than predictions made using an analytical model~\cite{Abd22JNR}.



The largest surface energy for BCC systems necessarily implies that other structures will be energetically favored below a given size. Using Eq.~\ref{eq:3}, the critical transition between BCC and phase $X$ is predicted to occur at 

\begin{equation}
N_c=\left[\frac{\beta(\mathrm{BCC})-\beta(\mathrm{X})}{\varepsilon_0(\mathrm{X})}\right]^3,\label{eq:4}
\end{equation}

since $\varepsilon_0 (\mathrm{BCC})=0$. Computed values are reported in Tab.~\ref{tab:2}. We find that FCC nanoparticles become more stable than BCC nanoparticles for $N\leq17$. This is dramatically smaller than previously reported sizes of $N_c=5660$ (Ref.~\onlinecite{Tom83PRB}) and $N_c=10470$ (Ref.~\onlinecite{Oh99JCP}). In addition, such a transition is not relevant here because our data also suggest that a disordered state is favored for $N\leq40$. It is well known that a non-crystalline molecular configuration should prevail for all metals at the smallest scales. Our computed threshold is close to a measured value of $N=30$, but this good agreement may be fortuitous given the large experimental uncertainty~\cite{Oh99JCP} and the limited set of disordered configurations considered in our simulations. 

To conclude this section, our DFT calculations unambiguously show that a BCC structure is favored for small tungsten nanoparticles of 1-1.6~nm. Nanoparticles with FCC or A15 structures are always higher in energies. When the number of atoms is lower than about 40, a disordered state becomes favorable. These results are at odds with a theoretical analysis~\cite{Tom83PRB} and X-ray diffraction experiments~\cite{Oh99JCP}. They are however in agreement with microscopy measurements~\cite{Sch14CCOM} and classical molecular dynamics calculations~\cite{Mar87JPCH,Che17JNR}. The main factor explaining the BCC stability over FCC despite the higher surface energy contribution, is the large bulk energy difference in favor of BCC. It is then obvious that the BCC structure will be favored for nanoparticles larger than those investigated here. Finally, it is important to keep in mind that we compute 0~K energy differences, and an evaluation of nanoparticle free energies would be needed for more definite conclusions. As a first hint, the bulk entropy difference between FCC and BCC is estimated to be approximately 1~$k_B$ per atom~\cite{Ozo09PRL}, thus at least one order of magnitude lower than the energy differences in Tab.~\ref{tab:1} at 300~K.  

\section{\label{sec:model}Semi-empirical modeling}

\begin{figure}[th!]
\begin{center}
\includegraphics[width=0.95\linewidth]{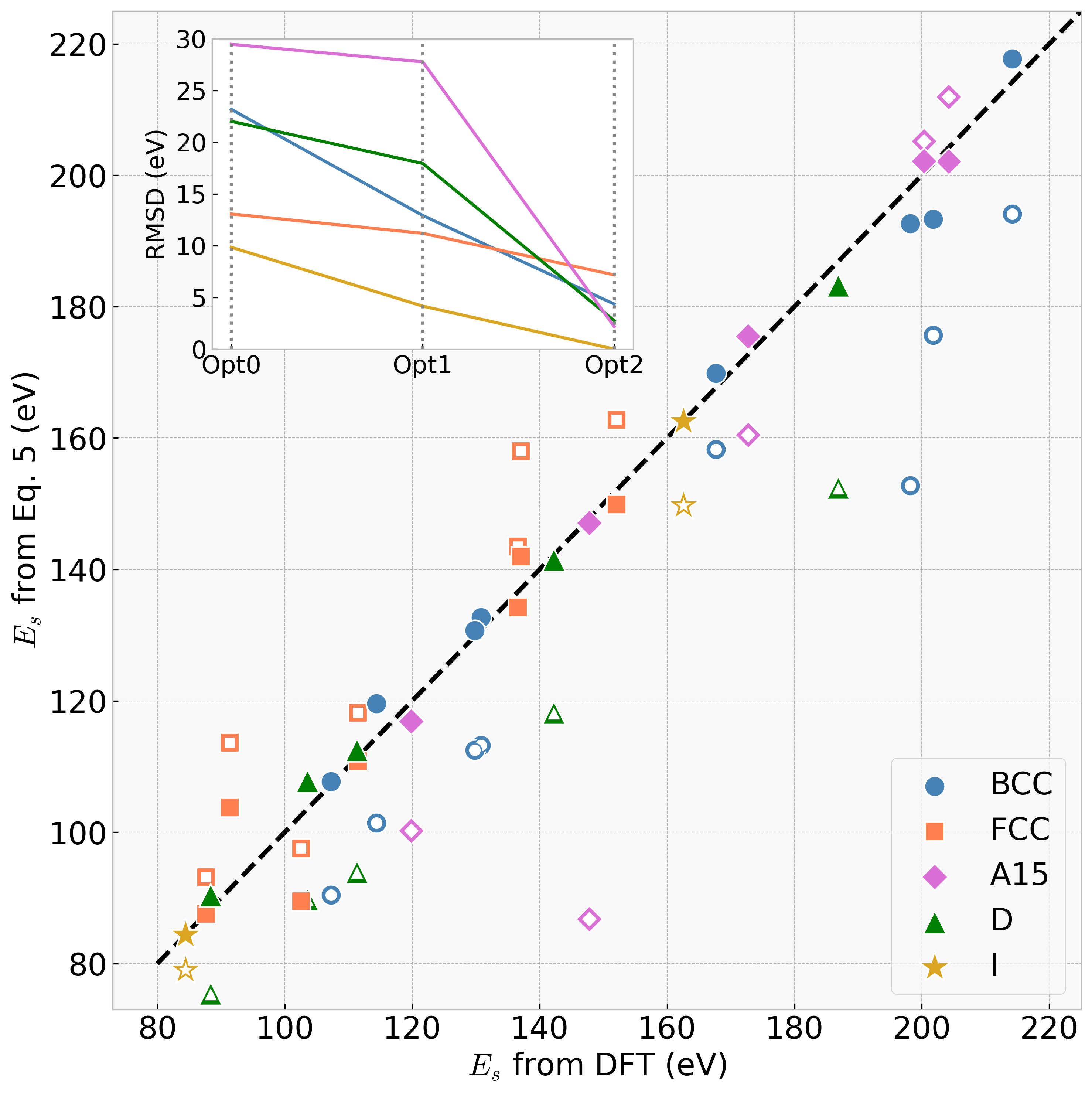}
\end{center}  
\caption{\label{fig:3} Surface energy $E_s$ computed with the Eq.~\ref{eq:t1} plotted against $E_s$ calculated by DFT, for all nanoparticles (BCC: blue circles; FCC, orange squares; A15: magenta diamonds; D: green triangles; I: golden stars). Open symbols show values obtained using the initial parameters as given in Ref.~\onlinecite{Tom83PRB}  and reported in Tab.~\ref{tab:3} (Opt0, see text for details), while filled symbols correspond to those obtained using optimized parameters (Opt2). The dashed black line marks the perfect agreement. The inset graph shows the root mean square deviations (RMSD) for each set (lines with same colors than the points in the main graph) and different optimization levels (Tab.~\ref{tab:3}).}
\end{figure}

\renewcommand{\arraystretch}{1.3}
\begin{table}
\caption{\label{tab:3}Root mean square deviations (RMSD, in eV) of $E_s$(DFT)-$E_s$(Eq.~\ref{eq:t1}) for each set and different optimization levels. Opt0: $E_s$(Eq.~\ref{eq:t1}) is calculated using the original parameters ($\lambda$,$\eta$) given in Ref.~\onlinecite{Tom83PRB}. Opt1: $\eta$ is optimized. Opt2: both $\lambda$ and $\eta$ are optimized.}
\begin{ruledtabular}
\begin{tabular}{clccccc}
Opt & & BCC & FCC & A15 & D & I \\ \hline
\parbox[t]{2mm}{\multirow{3}{*}{0}} 
& $\eta$ & 0.4 & 0.08 &  0.4   & 0.08 & 0.08\\ 
& $\lambda$ & 0.5 & 0.5 & 0.5 & 0.5 & 0.5\\ 
& RMSD & 23.17 & 13.06 & 29.46 & 22.01 & 9.84 \\ \hline
\parbox[t]{2mm}{\multirow{3}{*}{1}} 
& $\eta$ & 1.00 & 0.20 & 0.42 & 0.00 & 0.33 \\ 
& $\lambda$ & 0.5 & 0.5 & 0.5 & 0.5 & 0.5\\ 
& RMSD & 12.91 & 11.19 & 27.76 & 17.94 & 4.15 \\ \hline
\parbox[t]{2mm}{\multirow{3}{*}{2}} 
& $\eta$ & 1.00 & 0.20 & 0.77 & 0.04 & 0.12\\ 
& $\lambda$ & 0.546 & 0.461 & 0.579 & 0.619 & 0.540 \\ 
& RMSD & 4.35 & 7.17 & 2.20  & 2.76 & 0.00\\ 
\end{tabular}
\end{ruledtabular}
\end{table}

%

In this section, we focus on the semi-empirical model proposed by Tom{\'a}nek and co-workers~\cite{Tom83PRB}. We first aim at comparing and improving predictions from this model with our DFT calculated nanoparticle energies. In a second step, we try to understand why using this model leads to overestimated sizes for the BCC-FCC transition. 

We summarize the basics of this model in the first place. Using the second moment approximation, Tom{\'a}nek and co-workers proposed that $E_s$ in Eq.~\ref{eq:1} can be approximated by 

\begin{equation}
E_s=E_c\sum_{i=1}^{N_s}\left[(Z_i/Z_b)^{\lambda}-1\right],\label{eq:t1}
\end{equation}

where $N_s$ is the number of surface atoms and $\lambda=1/2$. $E_c$ is the cohesive energy of the bulk state. $Z_i$ is an effective coordination number including both first and second neighbors: $Z_i=Z_i^1+\eta Z_i^2$ where $Z_i^1$ and $Z_i^2$ refer to nearest neighbors and next-nearest neighbors, respectively. In their paper Tom{\'a}nek and co-workers reported that appropriate values for parameter $\eta$ are 0.08 for FCC and 0.4 for BCC~\cite{Tom83PRB}. They also defined surface atoms as those with $Z_i$ lower than 10. Finally, $Z_b$ is the coordination for bulk atoms, which depends on the atomic structure. 

We first compute $Z_i$ for all BCC and FCC nanoparticles studied in the present work, from the DFT relaxed configurations and following the rules mentioned above. Using $E_c(\mathrm{BCC})=-8.90$~eV/at~\cite{Kit76WIL} and $E_c(\mathrm{FCC})=-8.90+\varepsilon_0=-8.406$~eV/at, $E_s$ is calculated according to Eq.~\ref{eq:t1}. The results are plotted against DFT data in Fig.~\ref{fig:3}, and labeled as 'Opt0'. It appears that compared to DFT $E_s$ for BCC systems is systematically underestimated, whereas it is overestimated for FCC systems. The root mean square deviations (RMSD) are reported in Tab.~\ref{tab:3}. One can see that the error is significant, in particular for BCC nanoparticles.  

Nanoparticles with A15, disordered and icosahedral structures are not considered in Ref.~\onlinecite{Tom83PRB}, and we have to determine appropriate guess for model parameters. For disordered D and icosahedral I nanoparticles, we use the same parameters than for FCC ones. We know that $E_c(\mathrm{D})=-8.593$~eV/at and $E_c(\mathrm{I})=-8.467$~eV/at, using $\varepsilon_0$ from Tab.~\ref{tab:2}. In the A15 structure, there are 8 atoms in the elementary cell, with two different atomic environments. 2 of them have 12 neighbors at 2.83~\AA, and the remaining 6 have 2 neighbors at 2.53~\AA, 4 at 2.83~\AA, and 8 at 3.102~\AA. We assume that the first neighbors distance is lower than 2.95~\AA, i.e. $Z_i^1=7.5$ on average. Neighbors atoms at 3.102~\AA\ are considered second neighbors, i.e $Z_i^2=6$ on average. We use $\eta=0.4$ as initial guess by analogy with BCC, and $E_c(\mathrm{A15})=-8.807$~eV/at. In all cases, atoms are identified as belonging to the surface if their effective coordination number is lower than 10, as in Ref.~\onlinecite{Tom83PRB}. $E_s$ values computed with the model for A15, D and I nanoparticles are included in Fig.~\ref{fig:3}. In most cases they are underestimated compared to the DFT results. The RMSD for D and A15 are similar to the BCC value (Tab.\ref{tab:3}). Note that the low RMSD value for I is due to the limited size of the set (two nanoparticles).  

In order to improve the accuracy of the semi-empirical model, an optimization method is applied with $\eta$ as a variable parameter (Opt1). The new $\eta$ and RMSD values are shown in Tab.~\ref{tab:3}. The best improvements are obtained for BCC and I, but only a moderate RMSD reduction is observed for FCC, A15 and D nanoparticles. Finally, we allow both $\eta$ and $\lambda$ to vary during the optimization (Opt2). Final results are plotted in Fig.~\ref{fig:3}. A remarkable refinement is achieved for A15, D and BCC systems, and to a lesser extent for FCC. This is clearly demonstrated by the RMSD values which become lower than 3~eV for D and A15. Overall, we find that the model is highly sensitive to $\lambda$, and much less to $\eta$. Except for FCC, a $\lambda$ value slightly greater than 0.5 greatly increases the model accuracy.

\begin{figure}[th!]
\begin{center}
\includegraphics[width=0.95\linewidth]{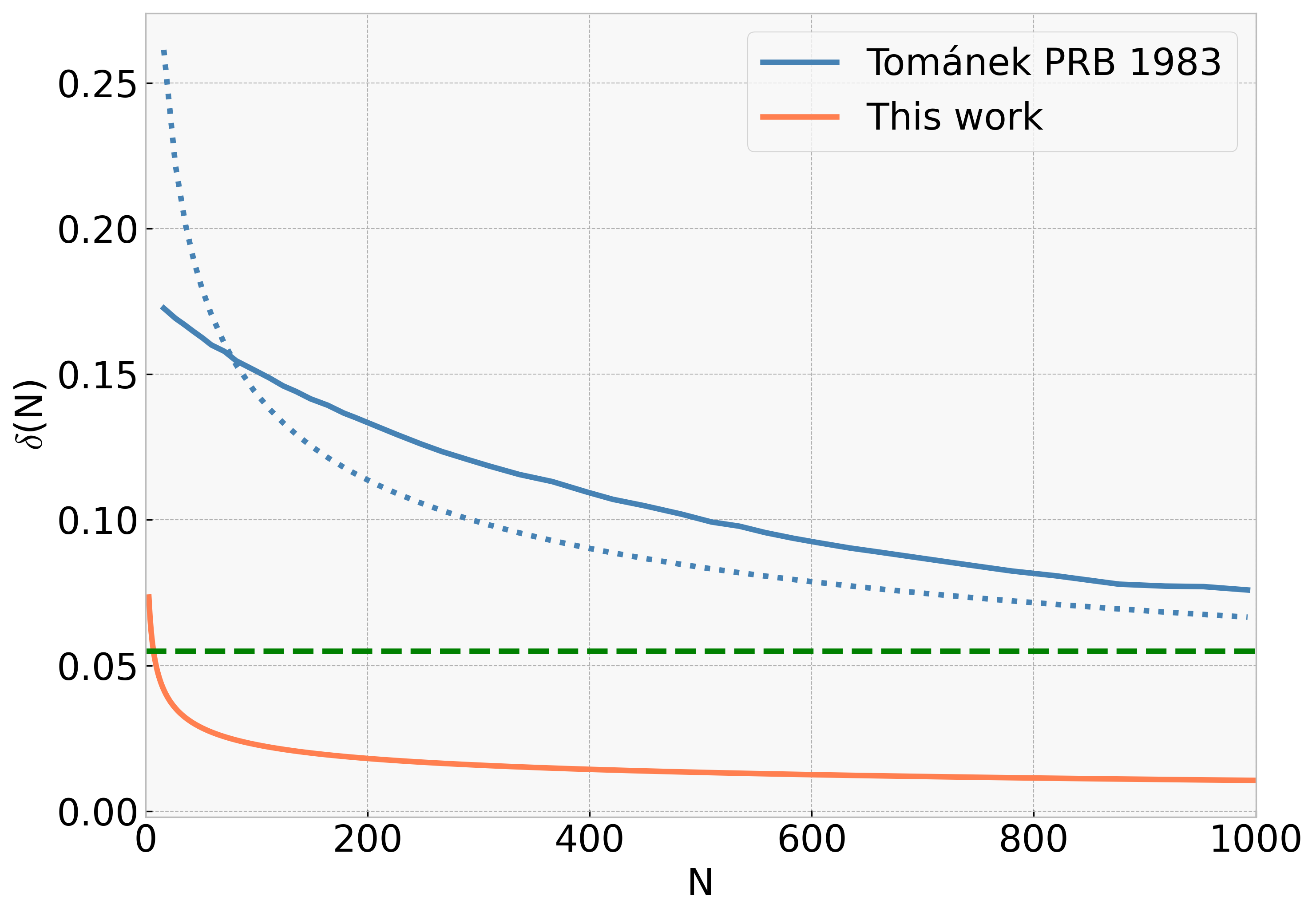}
\end{center}
\caption{\label{fig:4} $\delta(N)$ as a function of $N$, the number of atoms: data from Fig.~3 in Ref.~\onlinecite{Tom83PRB} (blue line), and computed from DFT data using Eq.~\ref{eq:t4} (orange line). The dotted blue line shows the best fit of $\delta(N)$ data in Ref.~\onlinecite{Tom83PRB} with the expression $\mu N^{-1/3}$. The dashed green line is positioned at $(E_c(\mathrm{BCC})-E_c(\mathrm{FCC}))/E_c(\mathrm{BCC})=0.0555$ (using $E_c(\mathrm{BCC})=-8.90$~eV and $E_c(\mathrm{FCC})=-8.406$~eV).   }
\end{figure}

Surface energies calculated using the original model in Ref.~\onlinecite{Tom83PRB} are therefore not very accurate, but it is not clear whether this is the main reason behind the difference in prediction for the BCC-FCC transition. In the same paper, the following expression is defined: 

\begin{equation}
\delta(N) = \frac{1}{N}\left[\sum_{\substack{i=1\\\mathrm{FCC}}}^{N_s}\left[(Z_i/Z_b)^{1/2}-1\right]-\sum_{\substack{i=1\\\mathrm{BCC}}}^{N_s}\left[(Z_i/Z_b)^{1/2}-1\right]\right]. \label{eq:t2}
\end{equation}

$\delta(N)$ depends only on the cluster geometry and is represented in the Fig.~3 of Ref.~\onlinecite{Tom83PRB} and reproduced in our Fig.~\ref{fig:4}. Tom{\'a}nek and co-workers proposed that the FCC-BCC transition occurs when the following condition is met:

\begin{equation}
 \delta(N) \stackrel{N=N_c}{=} \frac{E_c(\mathrm{BCC})-E_c(\mathrm{FCC})}{E_c(\mathrm{BCC})},  \label{eq:t2b}
\end{equation}

with the right-hand side a constant, plotted as a dashed line in Fig.~\ref{fig:4}. They found that $N_c=5660$ using their $\delta(N)$ values. 

We remark that $\delta(N)$ can also be computed from the $\beta$ values in Tab.~\ref{tab:2} determined from DFT results. In fact, using Eq.~\ref{eq:t1}, we can write




\begin{equation}
\delta(N)=\frac{1}{N}\left[\frac{E_s(FCC)}{E_c(\mathrm{FCC})}-\frac{E_s(BCC)}{E_c(\mathrm{BCC})}\right]. \label{eq:t3}
\end{equation}

Since $E_s = \beta N^{2/3}$, we finally obtain

\begin{equation}
\delta(N)=N^{-1/3}\left[\frac{\beta(\mathrm{FCC})}{E_c(\mathrm{FCC})}-\frac{\beta(\mathrm{BCC})}{E_c(\mathrm{BCC})}\right]. \label{eq:t4}
\end{equation}

Figure~\ref{fig:4} shows $\delta(N)$ computed with Eq.~\ref{eq:t4}, $\beta$ values in Tab.~\ref{tab:2}, $E_c(\mathrm{BCC})=-8.90$~eV and $E_c(\mathrm{FCC})=-8.406$~eV. There is clearly a large difference with $\delta(N)$ as given in Ref.~\onlinecite{Tom83PRB}, which likely explains the disagreement concerning the critical size of the BCC-FCC transition. With our data and using Eq.~\ref{eq:t2b}, we find a transition at $N_c=7$. The small difference with $N_c=17$ in Tab.~\ref{tab:2} can be explained by the fact that an additional approximation is made in Ref.~\onlinecite{Tom83PRB} to derive the Eq.~\ref{eq:t2b}.  


It is also noteworthy that the $\delta(N)$ curve provided in Ref.~\onlinecite{Tom83PRB} does not seem physically correct. In fact, it should mainly obey a $N^{-1/3}$ variation. The best fit of $\delta(N)$ with the expression $\mu N^{-1/3}$ is represented as a dotted blue line in Fig.~\ref{fig:4}, with $\mu=0.6646$. The agreement is obviously not satisfactory. This cast some doubts about the validity of these $\delta(N)$ data. In addition, according to Eq.~\ref{eq:t4} one can write
\begin{equation}
\mu=\frac{\beta(\mathrm{FCC})}{E_c(\mathrm{FCC})}-\frac{\beta(\mathrm{BCC})}{E_c(\mathrm{BCC})}.\label{eq:t5}
\end{equation}
This expression can be employed to compute $\beta(FCC)$ assuming that $\beta(BCC)=7.149$~eV, $E_c(\mathrm{BCC})=-8.90$~eV,  $E_c(\mathrm{FCC})=-8.406$~eV, and $\mu=0.6646$. We find $\beta(FCC)=1.166$~eV, which corresponds to $\gamma=0.601$~J~m$^{-2}$ (see the Appendix). Such a surface energy value is too low for W to be physically meaningful. This confirms that the main source of discrepancy between our calculations and predictions made in Ref.~\onlinecite{Tom83PRB} is the $\delta(N)$ curve. Since the latter is not a material dependent quantity, this unfortunately questions the validity of their predictions for other metals. 

\section{\label{sec:conclusions}Conclusions}

In this paper we report investigations on the structure and stability of tungsten nanoparticles, based on first-principles DFT calculations. In particular, various nanoparticles with BCC, FCC, A15, disordered and icosahedral structures, are considered. These models include 55 to 169 atoms, equivalent to sizes of about 1--1.6~nm. Our first conclusion is that BCC nanoparticles are the most stable energetically, followed by A15, disordered, icosahedral, and FCC in this order. Variations of nanoparticle energy as a function of size reveal a BCC-disordered transition at small sizes (at 40 atoms), and no BCC-FFC transition. It contradicts an earlier benchmark theoretical study on transition metal nanoparticles~\cite{Tom83PRB}. Our investigations suggest that the discrepancy could be explained by an erroneous description of the cluster geometry in this study. Finally, we also analyze the proposed semi-empirical model based on the second moment approximation, and improve its accuracy by adjusting the model parameters with respect to our DFT data. This will be useful for accurately calculating the energy of tungsten nanoparticles for all investigated structures, at much larger sizes and without the need of explicit atomistic calculations. 

As it stands, our calculations do not explain why A15 and FCC tungsten nanoparticles were observed~\cite{Iwa85SS,Oh99JCP}, and not only BCC ones~\cite{Sch14CCOM}. The most likely rationale is that these structures can be formed in specific growth conditions, with a possible influence of substrate, impurities, or surfactants. Although they are metastable compared to BCC, once formed they are probably stable at ambient conditions or during moderate annealings.   

In perspective to this study, in the light of the results presented here, it seems worthwhile to perform DFT calculations of the structure and stability of nanoparticles made of other transition metals, in particular those for which a BCC-FCC transition was predicted like Mo, Ta, Nb, Cr and V. 

\begin{acknowledgments}
Computer time for this study was partially provided by the MCIA
(Mésocentre de Calcul Intensif Aquitain). This work pertains to the
French Government program “Investissements d’Avenir” (EUR INTREE,
reference ANR-18-EURE-0010, and LABEX INTERACTIFS, reference
ANR-11-LABX-0017-01).
\end{acknowledgments}

\ \\
\section*{Author Declarations}

\subsection*{Conflict of interest}
The authors have no conflicts to disclose.

\subsection*{Author contributions}
\textbf{Laurent Pizzagalli:} Conceptualization (lead); Investigation (lead); Writing – original draft (lead); Writing – review \&
editing (equal). \textbf{Sandrine Brochard:} Conceptualization (supporting); Writing – review \& editing (equal). \textbf{Julien Durinck:} Conceptualization (supporting); Writing – review \& editing (equal). \textbf{Julien Godet:} Conceptualization (supporting); Writing – review \& editing (equal). 

\section*{Data Availability Statement}
The data that support the findings of this study are available from the corresponding author upon reasonable request.

\appendix*

\section{\label{sect:spheremod}Spherical nanoparticle model}

A general expression of the energy $E$ of a nanoparticle of $N$ atoms is 

\begin{equation}
E = N\varepsilon_0 + \gamma S(N),\label{eq:A1}
\end{equation}

with $S(N)$ the surface of the nanoparticle, $\gamma$ the surface energy, and $\varepsilon_0$ the energy per atom of the corresponding bulk phase. This is equivalent to Eq.~\ref{eq:1} if $E_s(N)=\gamma S(N)$. Assuming a spherical shape for the nanoparticle and a radius $r$, the nanoparticle surface is 

\begin{equation}
S = 4\pi r^2,\label{eq:A2}
\end{equation} 

and the nanoparticle volume is 

\begin{equation}
V= \frac{4\pi}{3}r^3=Nv_0,\label{eq:A3}
\end{equation}

with $v_0$ the volume of one atom. Combining Eqs.~\ref{eq:A2} and \ref{eq:A3}, we obtain

\begin{equation}
r = \left(\frac{3v_0N}{4\pi}\right)^{1/3},
\end{equation}

and 

\begin{equation}
S = 4\pi\left(\frac{3v_0}{4\pi}\right)^{2/3}N^{2/3}.
\end{equation}

Finally, Eq.~\ref{eq:A1} can be written

\begin{equation}
E = N\varepsilon_0 + 4\gamma\pi\left(\frac{3v_0}{4\pi}\right)^{2/3}N^{2/3},
\end{equation}

and  

\begin{equation}
E/N = \varepsilon_0 + 4\gamma\pi\left(\frac{3v_0}{4\pi}\right)^{2/3}N^{-1/3}.
\end{equation}

In comparison with Eq.~\ref{eq:3}, we finally get

\begin{equation}
\gamma = \frac{\beta}{4\pi}\left(\frac{4\pi}{3v_0}\right)^{2/3}.\label{eq:A4}
\end{equation}

$v_0$ is taken to be equal to the bulk atomic volume, assuming that the nanoparticle relaxation is small, and that $v_0$ is the same for all nanoparticle atoms. $v_0$ is then easily calculated for each crystalline structure using DFT determined $a_0$ (given in Tab.~\ref{tab:1}). For the disordered D and icosahedral I nanoparticles, we use the same $v_0$ than for the FCC structure.

\section*{References}
%

\end{document}